\documentclass{article}
\usepackage{epsfig}

\date{\today}

\usepackage{amsmath}

\newcommand{\ee}{\end{equation}}
\newcommand{\eea}{\end{eqnarray}}
\newcommand{\be}{\begin{equation}}
\newcommand{\bea}{\begin{eqnarray}}

\begin{document}

\title{Boson stars in SU(2) Yang-Mills-scalar field theories}
\author{{\large Yves Brihaye \footnote{yves.brihaye@umh.ac.be}}\\
\small{
Facult\'e des Sciences, Universit\'e de Mons-Hainaut,
B-7000 Mons, Belgium }\\
{ }\\
{\large Betti Hartmann \footnote{b.hartmann@iu-bremen.de}}\\
\small{
School of Engineering and Science, International University Bremen, 28725 Bremen, Germany
 }\\
{ }\\
{\large Eugen Radu\footnote{radu@heisenberg1.thphys.may.ie}}\\
\small{
Department of  Mathematical Physics,
National University of Ireland Maynooth, Ireland}}

\date{\today}

\maketitle
\begin{abstract}
We present new spherically symmetric solutions of an
 SU(2) Einstein-Yang-Mills model coupled to a doublet of scalar fields.
Sequences of asymptotically flat, Yang-Mills-boson star-type configurations
are constructed numerically by
considering an appropriate time-dependent ansatz for
the complex scalar field and a static, purely magnetic SU(2)-Yang-Mills potential.
Both nodeless as well as solutions with nodes of the scalar field and
gauge potential are considered.
We find that these solutions share many features with the
``pure'' boson stars.
\end{abstract}

\section{Introduction}

Although there is still no direct evidence for the existence of
scalar fields, there are many theoretical reasons that these
fields might play an important role in the
evolution and the structure of the Universe.
It is possible that a fraction of the bosonic dark matter has collapsed to form 
stellar type objects - boson stars.
Boson stars (BS) are localised, static configurations of
gravitationally bound zero temperature scalar particles, 
the complex scalar field possessing a harmonic
time dependence.
The study of  BS started with the work of Kaup \cite{Kaup} and
Ruffini and Bonazzalo  \cite{Ruffini:1969qy}, who
found asymptotically flat, spherically symmetric equilibrium solutions
of the Einstein-Klein-Gordon equations.
These configurations are ``macroscopic quantum states'' and are only prevented
from collapsing
gravitationally by the Heisenberg uncertainty principle.

The BS share many features with their fermionic counterparts, presenting however
 many interesting differences.
For example, BS also exhibit a critical mass and critical particle number.
Later work considered the self-interacting case 
\cite{Mielke:1980sa,Colpi:1986ye,Schunck:1999zu}
or a non-minimal coupling of the scalar field to gravity \cite{vanderBij:1987gi}.
Boson stars in the presence of a dilaton or an axidilaton have
also been studied by various authors \cite{Gradwohl:1989px},
as well as boson-fermion stars \cite{Henriques:1989ez}.

All these models have demonstrated the same characteristic: new interactions tend to
increase the critical values
of mass and particle number, although the particular values are very model dependent.
The stability against perturbations
around the equilibrium state has been discussed also  by a number of authors
\cite{Gleiser:1988rq}-\cite{Kusmartsev:cr}.
An extensive review of the boson star properties is given in \cite{Jetzer:1992jr},
\cite{Mielke:1997re} and more recently in \cite{Shunck:2003vk}.

Jetzer and van der Bij extended the BS model to include the coupling with a $U(1)$
gauge group \cite{Jetzer:1989av}.
Here, the scalar field regularizes the central singularity of the
Reissner-Nordstr\"om solution, which otherwise
would necessarily be present in the pure Einstein-Maxwell theory.
A natural generalization of these charged BS
configurations is to consider a larger gauge group.
In this case, there are regular configurations even in the absence of a
scalar field, as proven by the Bartnik-McKinnon (BM) static, time-independent solution
of the coupled SU(2) Einstein-Yang-Mills (EYM) equations \cite{bm}
(see \cite{Volkov:1998cc} for a general discussion of the properties of this type of solutions).

In this paper, we extend the analysis of  \cite{Jetzer:1989av} to include a
complex doublet of scalar fields coupled to an SU(2) non-abelian gauge field
\footnote {Asymptotically anti de-Sitter 
q-stars interacting with a nonabelian field have been studied
recently in \cite{Prikas:2004fe}. However, the properties of these notopological solitons
are rather different as compared with the solutions discussed in this paper.}.
In the static case,  and for a symmetry breaking scalar  potential, this system admits
sphaleron solutions, describing the top of
the potential-energy barrier separating
gauge-inequivalent classical vacua \cite{Klinkhamer:1984di}.
The BM configurations are recovered in the limit of vanishing
scalar field \cite{greene}.

Here we present both numerical and analytical arguments
for the existence of a new type of solution of the coupled EYM-scalar field equations,
which combines the basic properties of both BS and BM models.
For any value of the scalar field at the origin,
the BS solutions can be generalized to include
a BM particle inside.
Different from the sphaleron case, however, the magnitude of the
scalar field is zero at infinity.

The paper is structured as follows:  in the next Section   we present the general
framework and analyse the field equations and boundary conditions.
In Section 3 we present our numerical results.
We conclude with Section 4, where our results are summarized.

\section{General framework and equations of motion}
\subsection{Basic ansatz}
Our study of the EYM-scalar field system is based upon the action
\begin{equation}
\label{lag0}
S=\int d^{4}x\sqrt{-g_m}\left[\frac{1}{16\pi G}\mathcal{R}
-(D_{\mu}\Phi)^{\dagger}(D^{\mu}\Phi)-V(\Phi)
-\frac{1}{4} F_{\mu \nu}^a F^{a \mu \nu}\right],
\end{equation}
where, following the standard model, we take the scalar field 
$\Phi$ to be a complex doublet.  
$G$ is  the  gravitational constant and $g_m$ is the determinant of the metric
tensor.

Here   $D_{\mu}$ is  the  usual  gauge-covariant
derivative
\begin{equation}
 D_{\mu} \Phi = \partial_{\mu} \Phi + g (A_{\mu}^a \sigma_a) \Phi  \ ,
\end{equation}
while
\begin{equation}
F_{\mu \nu}^a = \partial_{\mu} A_{\nu}^a
                    - \partial_{\nu} A_{\mu}^a
        + g \epsilon_{abc} A_{\mu}^b A_{\nu}^c
\end{equation}
is the SU(2) field strength tensor and $g$ is the gauge coupling constant.

In this letter we will restrict our analyse to the  case
$V(\Phi)=\mu^2 \Phi^{\dagger}\Phi$,
where $\mu$ is the scalar field mass, without including
a scalar self-interaction term.
As found in \cite{Colpi:1986ye},
although the inclusion of
a $\lambda |\Phi|^4$ term
drastically changes the value of the maximum mass and the corresponding
critical central density of the boson star solutions,
the qualitative features are essentially similar to the non-self interaction case.

The action is invariant under a global phase rotation
$\Phi \to \Phi e^{-i\alpha}$ which
implies the existence of a conserved current
\begin{eqnarray}
\label{J}
J^{\mu }=i g^{\mu \nu}
\left ( (D_{\nu}\Phi^{\dagger}) \Phi - (D_{\nu}\Phi) \Phi^{\dagger} \right) \ .
\end{eqnarray}
This gives an associated conserved charge, 
namely, the number of scalar particles:
\begin{eqnarray}
N=\int d^{3} x \sqrt{-g} J^t \ .
\end{eqnarray}

Since we assume spherical symmetry, it is convenient
to use the metric in Schwarzschild-like coordinates:
\begin{equation} \label{metric}
ds^{2}=\frac{dr^{2}}{B(r)}+r^2(d\theta^{2}+\sin^{2}\theta d\varphi^{2})-
\sigma^2(r) B(r)dt^{2} \ ,
\end{equation}
where $B(r)= 1-2m(r)/r$. $m(r)$ may be interpreted
as the total  mass-energy within the radius $r$; 
its asymptotic value gives the total ADM mass
of the solutions.

Similar to the pure BS case, we can define a configuration radius
\begin{eqnarray}
\label{R}
R=\frac{1}{N} \int~d^{3} x~\sqrt{-g} r J^t \ .
\end{eqnarray}

We consider only static gauge fields, and for simplicity we assume $A_t^a=0$.
The most general spherically symmetric SU(2) ansatz is parametrized by
\begin{equation}
A_i^a = \frac{1-W(r)}{gr} \epsilon _{aij} \hat x_j
       + \frac{\tilde{W}(r)}{gr} (\delta_{ia} - \hat x_i \hat x_a)
       + \frac{a(r)}{gr} \hat x_i \hat x_a   \ , 
\end{equation}
while for the scalar field we choose
\begin{equation}
 \Phi =
e^{-i \tilde{\omega} t} \bigl[ \phi(r) + i K(r) (\hat x^a \sigma_a) \bigr]
{0 \choose 1}
\ . \end{equation}

It is well known that the Ansatz for the matter fields
is plagued with a residual gauge symmetry. Along with \cite{akiba},
we  fix the gauge  by imposing the axial gauge
\be
      x^i A_i = 0 \ \ \  \Rightarrow ~~~a(r) = 0 \ \ .
\ee

In this paper we consider the consistent truncation of the full model
$\tilde{W}(r) = K(r) = 0$, which corresponds to imposing 
the symmetry of the fields under the parity operator. Note that the original
sphaleron solution of Klinkhamer and Manton \cite{Klinkhamer:1984di}
as well as the gravitating generalizations \cite{greene}
have been constructed in this reduced ansatz. 
However, solutions with $\tilde{W}(r)\neq 0$, $K(r)\neq 0$
(so-called ``bi-sphalerons'') have been constructed in \cite{bk,yaffe}.
The general spherically symmetric 
configurations have also been used in \cite{akiba}.
 
\subsection{Field equations and boundary conditions}
Within this ansatz, the classical equations of motion 
can  be derived from the following
2-dimensional action:
\begin{eqnarray}
\label{action}
S = \int dt \ dr ~\sigma  \left[\frac{m'}{4 \pi G}-
\frac{1}{g^2}\bigg(BW'^2+\frac{(W^2-1)^2}{2r^2}\bigg)-
\bigg(Br^2 \phi'^2+\frac{1}{2}\phi^2(W-1)^2+\mu^2 r^2\phi^2
-\frac{{\tilde{\omega}}^2\phi^2 r^2}{B \sigma^2}\bigg) \right],
\end{eqnarray}
where the prime denotes the derivative with respect to $r$.

The field equations  reduce to the
following system of four  non-linear differential equations:
\begin{eqnarray}
\label{equations}
\nonumber
m'&=& 4 \pi G \bigg(
\frac{1}{g^2}\left( B W'^2+\frac{(W^2-1)^2}{2r^2} \right)+
 Br^2 \phi'^2+\frac{1}{2}\phi^2(W-1)^2+\mu^2 r^2\phi^2
+\frac{\phi^2 r^2 \tilde{\omega}^2}{B \sigma^2}\bigg) \ ,
\\
\label{eqs}
\sigma'&=&\frac{8\pi G \sigma}{r} \left
(\frac{W'^2}{g^2}+ r^2 \phi'^2+\frac{\phi^2 r^2\tilde{\omega}^2}{ \sigma^2 B^2}\right) \ ,
\\
\nonumber
(\sigma B W')'&=&\frac{\sigma W}{r^2}(W^2-1)+\frac{1}{2}g^2\sigma \phi^2 (W-1) \ ,
\\
\nonumber
(\sigma B r^2 \phi')'&=&\frac{\sigma \phi}{2}(W-1)^2
+ \mu^2\sigma r^2 \phi-\frac{\phi r^2\tilde{\omega}^2}{\sigma B} \ .
\end{eqnarray}

The regularity of the solution at the origin, the finiteness
of the ADM mass   and the requirement
that the metric (\ref{metric})
approaches the Minkowski metric for $r\rightarrow \infty$
lead to definite boundary conditions  for the four
functions $m$, $\sigma$, $W$,  $\phi$.
As far as the metric functions are concerned we have to impose
\be
m(0) = 0 \ \ ,~~\sigma(\infty) = 1  \ .
\ee
The remaining functions have to obey
\begin{eqnarray}
W(0) = 1, ~~\phi'(0) = 0 \ ,
~~
W(\infty) = (-1)^p,~~\phi(\infty) = 0  \ ,
\end{eqnarray}
(where $p$ is a positive integer), which are the usual boundary
conditions for the BS and BM configurations,
respectively.
Thus, for finite energy solutions, the term
$\tilde{\omega}^2 \phi^2$ in (\ref{action}) forces $\phi$ to tend to zero at infinity,
which excludes a ``boson star-sphaleron'' hybrid model
with a scalar field potential.

The field equations imply
the following behaviour for $r \to 0$ in terms of three parameters
$(b,\phi_0,\sigma_0)$:
\begin{eqnarray}
\label{exp-origin}
\phi(r)&=&\phi_0+\frac{\phi_0}{6}(\mu^2 -\tilde{\omega}^2 \sigma_0^2) r^2+O(r^3) \ ,
~~~W(r)=1-b r^2+O(r^4) \ ,
\\
\nonumber
m(r)&= & \frac{1}{12\pi G g^2}\left( 6 b^2+\phi_0^2 g^2
(\mu^2-3\tilde{\omega}^2\sigma_0^2) \right) r^{3}+O(r^4) \ ,
~~~
\sigma(r)=\sigma_0+\left( \frac{b^2}{\pi G g^2}
+ \frac{\phi_0^2\tilde{\omega}^2\sigma_0^3}{4\pi G } \right)r^2+O(r^3) \ .
\end{eqnarray}
A similar analysis for  $r \to \infty$ reveals that
the leading order asymptotic behaviour of the metric functions is 
determined by the gauge field only,
while  the scalar field asymptotic form is similar to the pure BS case:
\begin{eqnarray}
\label{exp-inf1}
\phi(r)&=& \phi_1r^{\Delta}
e^{-\sqrt{\mu^2-\tilde{\omega}^2}r}+\dots \ \ ,
\ \ {\rm with} \ \ \Delta\equiv -1+M\frac{2\tilde{\omega}^2-\mu^2}{\sqrt{\mu^2-\tilde{\omega}^2}} \ ,
\\
\nonumber
W(r)&=&(-1)^p-\frac{w_1}{r}+\dots \ , \ \
m(r)= M-\frac{4\pi G}{g^2}\frac{w_1^2}{r^3}+\dots \ , \ \
\sigma=1-\frac{2\pi G}{g^2}\frac{w_1^2}{2r^4}+\dots \ .
\end{eqnarray}

To perform numerical computations and order-of-magnitude estimations,
it is useful to have a new set of dimensionless variables.
This is obtained by using the following rescaling
\begin{eqnarray}
\label{resc}
r \to r\sqrt{4 \pi G}/g  \  ,  \ \phi(r) \to \phi/\sqrt{8 \pi G} \ , \
m(r) \to m(r)\sqrt{4 \pi G}/g \ .
\end{eqnarray}

The system of differential equations (\ref{equations}) with the above boundary
conditions then depends on only two coupling constant, namely:
 \begin{equation}
 \alpha=\mu \frac{\sqrt{4 \pi G}}{g}=\frac{\sqrt{4\pi}}{g}\frac{\mu}{M_{Pl}} \ , \ 
\omega=\tilde{\omega} \frac{\sqrt{4\pi G}}{g}=\frac{\sqrt{4\pi}}{g}\frac{\tilde{\omega}}{M_{Pl}}
\end{equation}
where $M_{Pl}=\sqrt{G}$ is the Planck mass scale.

Note that
the factor $\omega$ could be absorbed into the definition
of the metric function $\sigma$ and
this would  change the boundary
conditions for $\sigma$, with $\sigma (\infty) \neq 1$. 
The system of differential equations (\ref{equations}) would then depend only on $\alpha$.

We have however kept two coupling constants,
$\alpha$ and $\omega$. In our numerical computations, 
we have used $\alpha$ and $\phi(0)$ 
as free parameters and obtained $\omega$ as function of $\phi(0)$.

\section{Numerical results}
The equations of motion (\ref{eqs})  have been solved for a range of values of
the scalar field at the origin, $\phi(0)$, and several values of $\alpha$.
As expected, these solutions have many features in
common with the pure BS and BM solutions;
they also present new features that
we will  point out in the discussion.

For each choice of $\phi_0$, the field equations
have a solution with the right asymptotics for $r \to \infty$
only when $(\sigma(0),~W''(0)\sim b)$ in
the expansion at the origin  (\ref{exp-origin})
take on certain values.
The value of the shooting parameter $b$ decreases with increasing  $\phi_0$.
For the same value of $\phi_0$, different values of $\sigma_0$
corresponds to different numbers of nodes of the scalar
field.

In the following, we denote by $p$ the number of nodes of the gauge field
functions and by $n$ the number of nodes of the scalar field function.
For the considered configurations we could not find non-trivial nodeless
gauge field function. Thus $p=0$ implies $W(r)\equiv 1$.
 \subsection{Known configurations: Bartnik-McKinnon and
``pure boson'' star solutions}
 If we set the scalar field to zero,
 the corresponding equation is trivially satisfied and we are left
 with the spherically symmetric Einstein-Yang-Mills system, whose
 solutions are the BM configurations \cite{bm}.
 They consist of a discrete family of smooth solutions uniquely characterized
 by the number of nodes $p$ of the function $W(r)$, with $p \geq 1$.
 For these solutions
 the metric function $B(r)$ possesses a minimum at some
 finite value of $r$. 

\newpage
\setlength{\unitlength}{1cm}

\begin{picture}(18,7)
\label{fig1}
\centering
\put(2,0.0){\epsfig{file=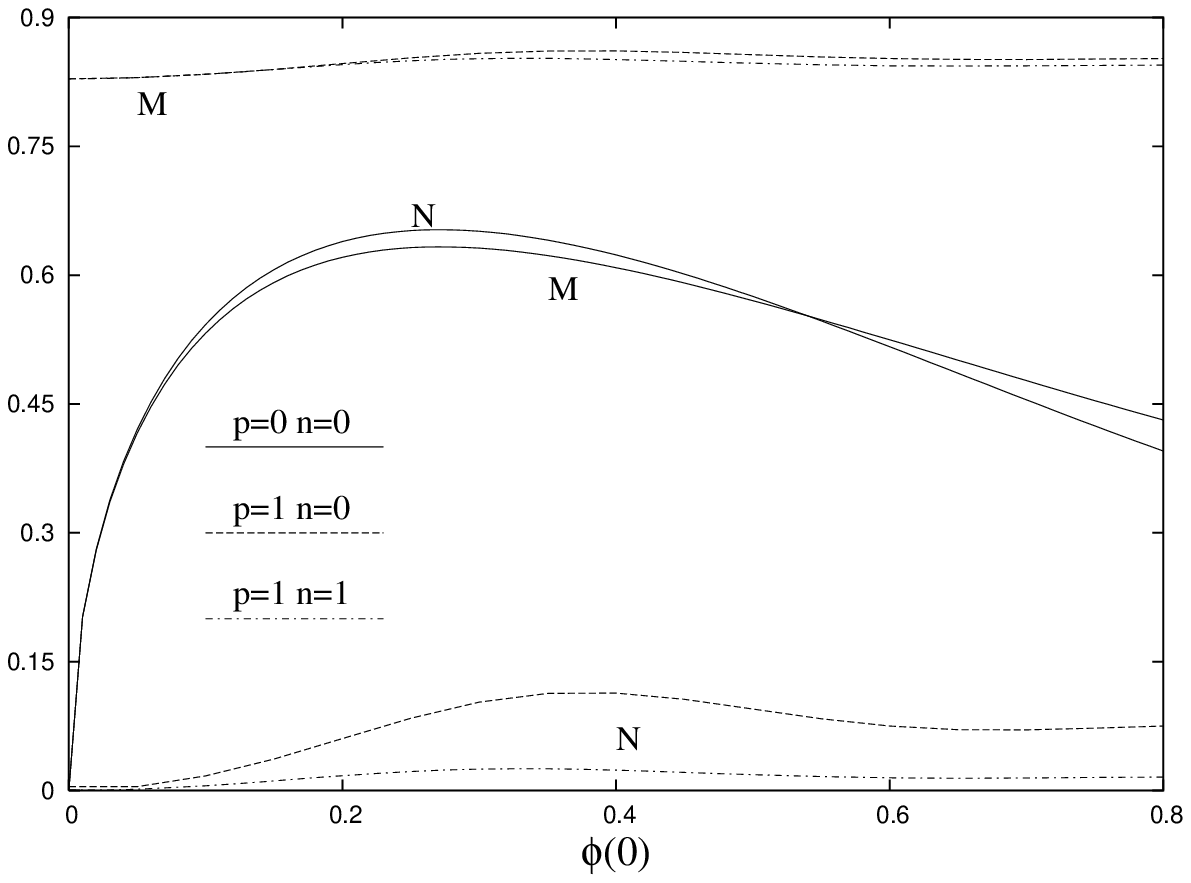,width=11cm}}
\end{picture}
\begin{picture}(19,8.)
\centering
\put(2.6,0.0){\epsfig{file=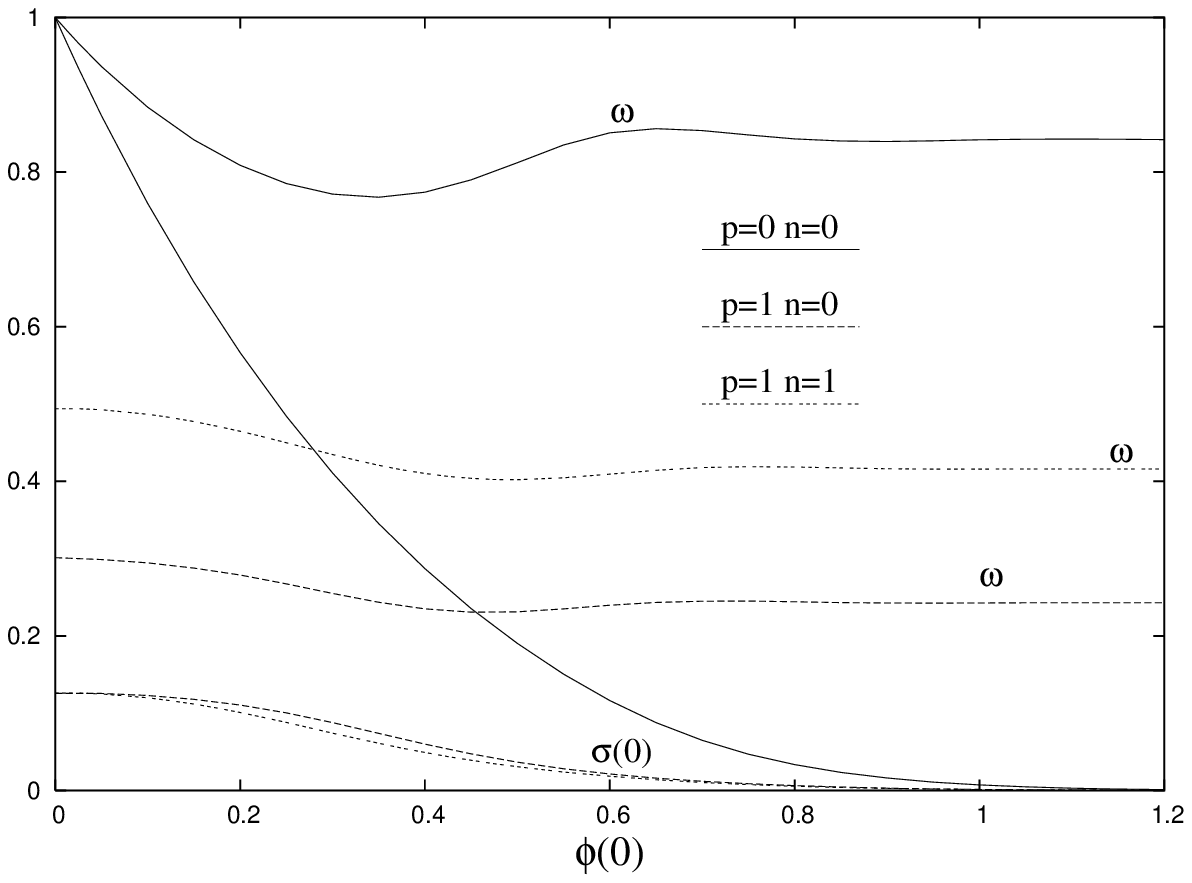,width=11cm}}
\end{picture}
\\
\\
{\small {\bf Figure 1.} The mass-parameter $M$, the particle number $N$,
the value of the metric function
$\sigma(r)$ at the origin, $\sigma(0)$, and the parameter $\omega$
 are represented as a function of the value of the scalar field at the origin
 $\phi(0)$ for ``pure''
boson stars $(p=0,~n=0)$ and $p=1$, $n=0$, $1$ Yang-Mills-boson star configurations.}
\\
\\
The metric
 function $\sigma(r)$ increases monotonically from $r=0$ (with $0<\sigma(0)<1$)
 to $r=\infty$ with $\sigma(\infty)=1$.
 Boson-star solutions discussed in this paper can be regarded as 
 deformations (i.e. with $\phi(r)\neq 0$)
 of each solution of the BM sequence.

In the case when the Yang-Mills field is  set to its vacuum value
the  parameter $\alpha$ can
 further be rescaled in the radial variable, and
 the equations redce to those of the ``pure'' boson-star.
 The scalar field can be gradually deformed by setting
 $\phi(0)=\phi_0 > 0$ and the corresponding solution can be constructed
 numerically \cite{Kaup, Ruffini:1969qy}.
 The configurations can be characterized namely by their mass ($m(\infty)$)
 the values $\phi(0)$, $\sigma(0)$ and $B_m$ (i.e. the minimal value of
 the metric function $B(r)$) and by the value  $\omega$.

  The behaviour of $M$ and $N$ as a function of $\phi(0)$ is well known
in this case; the  mass and
particle number rise with increasing $\phi(0)$ to
the maximum values $M_{max}=0.633$ and $N_{max}=0.653$ (see Figure 1).
Then $M$ and $N$ decrease, oscillate a bit and approach 
asymptotically a constant value.

\newpage
\setlength{\unitlength}{1cm}

\begin{picture}(18,7)
\label{fig2}
\centering
\put(2,0.0){\epsfig{file=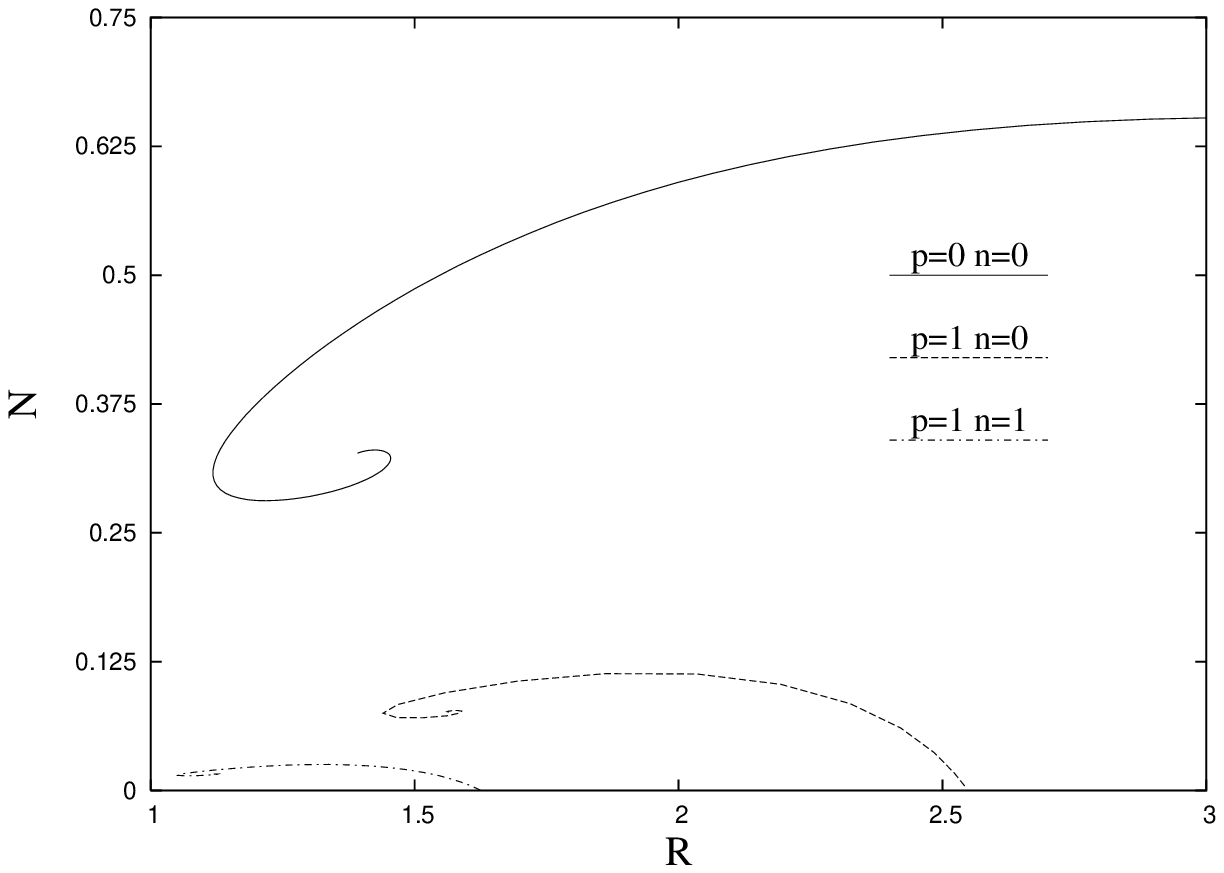,width=11cm}}
\end{picture}
\\
\\
{\small {\bf Figure 2.} The particle number $N$ is represented
as a function of the effective radius $R$ for the same configurations as in Figure 1.}
\\
\\

When the value of the scalar field at the origin $\phi(0)$ becomes
 large, the value $\sigma(0)$ decreases and tends asymptotically to zero.
However, a new phenomenon appears in this limit. Namely
 the metric function
$B(r)$ develops a second minimum between $r=0$ and the global minimum.
This will be discussed in more detail in
the context of YM-boson stars.

Figure 1 also illustrates the fact that BS solutions
 exist only on a definite interval of $\omega$
  and that on some subinterval ($0.76 \geq \omega \geq 0.85$)
 several solutions with the same $\omega$ can exist.
 Here we have examined only the case when the scalar field
 possesses no nodes (i.e. $n=0$), but solutions exist for which the
 scalar field function possesses one or more zeros, i.e. $n\neq 0$.
\subsection{Yang-Mills-Boson-star: $p=1,~2,~3$}
We have constructed 
(again by imposing a fixed value for $\phi(0)=\phi_0 > 0$)
families of solutions with $n=0$, $1$, $2$,
respectively $3$ nodes of the scalar field
function and $p=1$, $2$, $3$ nodes of the gauge field
function for varying $\alpha$.
We first discuss the solution for a generic value of the
mass parameter $\alpha$, namely $\alpha=1$. 
 The results  turn out to be qualitatively the same
 for other values of  this parameter.
A discussion of the domain of existence of solution
in the parameter $\alpha$ is presented
at the end of this section.

The data characterizing the $n=0$ and the $n=1$ solutions
 is shown in Figures 1, 2 .
 For the case $n=2$ the pattern remains qualitatively the same.
 These graphs present the same quantities as
in the case of the ``pure'' boson stars.

The main difference is that the $p$-th
 BM solution is approached in the limit $\phi(0) \rightarrow 0$.
 Accordingly, the ADM-mass does not vanish in this limit, but is equal
to that of the corresponding BM solution.
 The particle number
behaves like $\phi(0)^2$ and as
 a consequence the star radius stays finite.
Excluding this latter point, note that the behaviour of $M$ and $N$ as 
a function of $\phi(0)$ is  similar
to that of the ``pure'' boson stars.

 We see also that the mass and the parameter  $\omega$
 depend only weakly on $\phi(0)$
 (this holds true also for $B_m$ which is not shown
 in the Figure), while $\sigma(0)$ decreases
 monotonically for increasing $\phi(0)$.
 As in the case without a non-abelian field, 
we see that the particle number and the mass
have their extrema, in particular their maximum, 
at the same value of the central density (characterized by $\phi(0)$),
decreasing monotonically to a constant asymptotic value.
The behaviour of the particle number (or mass) in dependence on the radius
is similar to the ``pure'' BS case (see Figure 2). 
In both cases
the
particle number versus radius curve shows a   

\newpage
\setlength{\unitlength}{1cm}

\begin{picture}(18,7)
\label{fig3}
\centering
\put(2,0.0){\epsfig{file=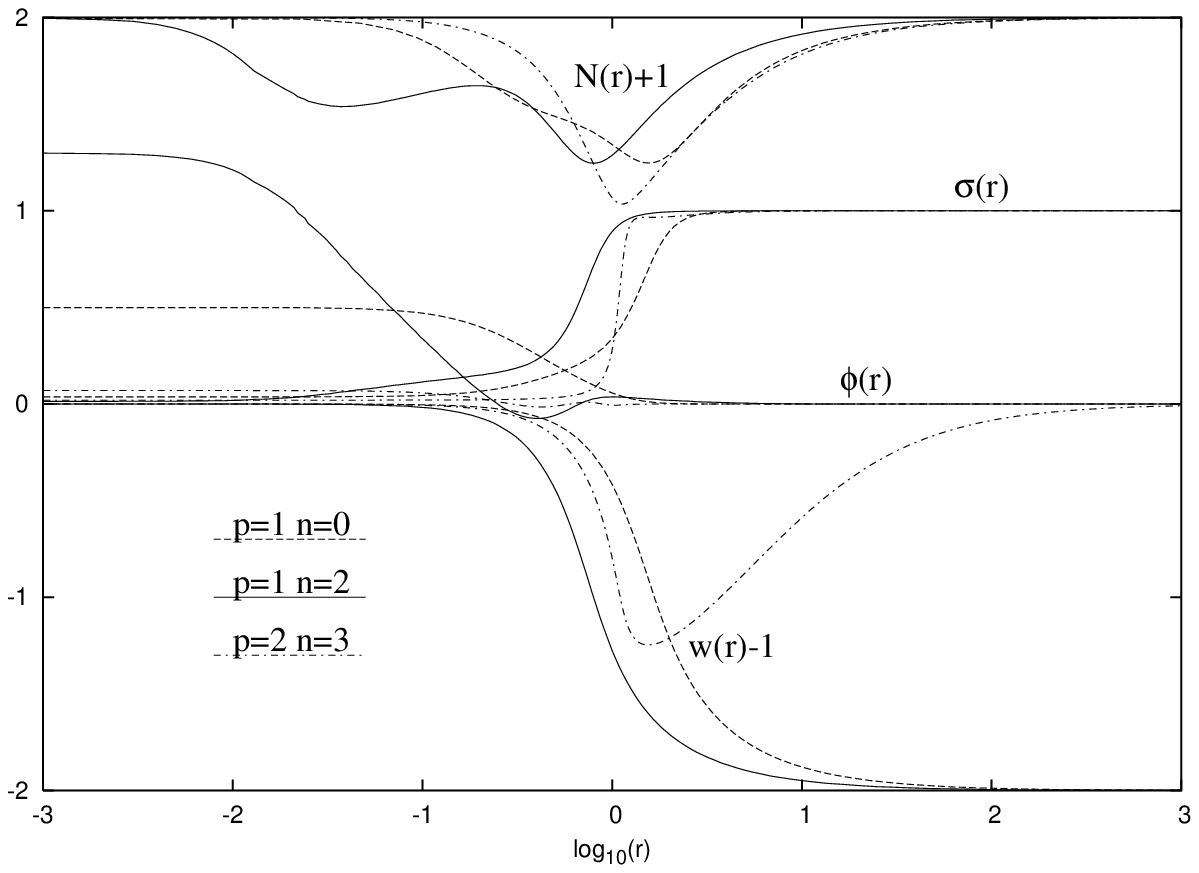,width=11cm}}
\end{picture}
\\
\\
{\small {\bf Figure 3.} The profiles of typical Yang-Mills-boson stars  with
$\alpha=1$
are plotted for several values of $p,~n$.}
\\
\\
counterclockwise inspiralling
at the critical
points corresponding to a maximum of $N$.
  Another point which is worth pointing
 out is that the various solutions exist only on definite
 intervals of the parameter $\omega$ and that
 $\omega$ is not a monotonic function
 of $\phi(0)$. This parameter converges to a finite value
 (depending on $n,p$) for $\phi(0) \rightarrow 0$.
Note that the values of $\omega$ for a fixed value of $\phi(0)$ 
are smaller in the case of Yang-Mills boson stars as compared to the
``pure'' boson stars and for the Yang-Mills boson stars themselves
smaller for $n=0$ as compared to $n=1$.

 Another feature which the Yang-Mills boson star 
 shares with the ``pure'' boson
star  is the fact that,
 when the scalar field's value at the origin becomes large enough,
 the metric function $B(r)$ develops a second local minimum, situated
 inside the global minimum.
 This is illustrated in Figure 3, where the profiles of the functions
 $B(r)$, $\sigma(r)$, $W(r)$, $\phi(r)$ are shown for
three different combinations of $p$ and $n$.
Note that for $p=1$ and $n=2$ the mentioned
second minimum, which we associate with a  
shell-like structure of the solutions, appears.

 We remark here that our numerical analysis was stopped due to numerical
difficulties related to the smallness of $\phi(0)$ and we can thus not make
any conclusive statement about whether the second minimum becomes the global
minimum if $\phi(0)$ is small enough.

We have also managed to construct solutions with two nodes 
in the gauge field,
i.e. $p=2$ and $n$ nodes in the scalar field. The profiles of the
functions $N(r)$, $\sigma(r)$, $W(r)$ and $\phi(r)$ for the  $p=2$ and
$n=3$ solution are shown in Figure 3. The qualitative behaviour
of $N$ and $M$ is the same as for $p=1$.
The most sensitive parameter to the number of nodes $p$ is the 
frequency $\omega$. Choosing $\phi(0)$ rather small (e.g. $\phi(0)=0.1$)
and concentrating on the $n=0$ solution,
we observe that $\omega \sim 0.3$   for $p=1$
and  $\omega \sim 0.9$   for $p=2$. This has important consequence
on the domain of existence of solutions in the $\alpha$-$\phi(0)$-plane.

Let us finally discuss this domain of existence. The precise
determination of this domain is a huge task, which we don't aim at in this
letter, however, we can make some statements about the qualitative behaviour.
We first assume that $\phi(0)$ is fixed (say $0 < \phi(0) < 1.5$)
and $\alpha$ is varied. 
Our numerical analysis suggests that the
parameter $\omega$ decreases considerably when $\alpha$ decreases.
A consequence of this is that the quantity 
$\sqrt{\alpha^2 - \omega^2}$, which appears in the asymptotic analysis
(see (\ref{exp-inf1})) becomes imaginary for 
$\alpha < \alpha_{cr}$ such that
exponentially localized solutions cease to exist for such values.
For $p=1$, we find numerically
$\alpha_{cr}\approx 0.16$ for $\phi(0) = 0.1$  and
 $\alpha_{cr}\approx 0.12$ for $\phi(0) = 0.5$.
For $p=2$ the critical value of $\alpha$ turns out to be larger,
e.g. $\alpha_{cr}\sim 0.85$ for $\phi(0)=0.1$.

We could hardly find any relevant critical phenomenon
limiting the
solution pattern when increasing $\alpha$. For instance
we checked that both the metric and the scalar field
stay perfectly regular up to $\alpha \sim 10.0$.

For fixed $\alpha$ and increasing $\phi(0)$ we have found
that the value of the metric function $\sigma$ at the origin,
$\sigma(0)$, becomes very small for large $\phi(0)$, while the minimum
of the function $B(r)$ stays finite. This suggests that the
solution becomes singular at the origin for large enough $\phi(0)$.
However, because the numerical calculations become very difficult for large $\phi(0)$, 
we
 refrain from making any further conclusions here.

\section{Conclusions and further remarks}
In this paper we have investigated
a new type of configuration combining the basic properties of
two well known solutions:
boson stars for a time-dependent complex scalar field coupled to
gravity on the one hand and Bartnik-McKinnon
particle-like solutions of the SU(2) Einstein-Yang-Mills system
 on the other hand.

The properties of these Yang-Mills boson star solutions 
differ considerably from their 
abelian counterparts discussed in \cite{Jetzer:1989av}.
Note also that these solutions represent
 the simplest examples of Yang-Mills boson star configurations.
We expect
that static, respectively rotating 
axially symmetric solutions  as well as dilaton generalizations
may exist as well.

The Einstein-Yang-Mills system possesses also black hole solutions,
so-called ``coloured'' black holes, which are counterexamples 
to the no-hair conjecture \cite{Volkov:1998cc}.
There are no black hole analogues to 
the ``pure'' BS configurations \cite{Pena:1997cy} though.
However, the introduction of an SU(2) gauge field invalidates the arguments
presented in \cite{Pena:1997cy}. The existence of black hole counterparts
of the configurations discussed in this paper is thus an open question.

An important physical question when discussing self-gravitating 
configurations is whether these solutions are stable.
The stability of the Yang-Mills boson stars can be studied in general
by considering a perturbation about the classical solutions and solving
the linearized equations. This leads
to a system of coupled Schr\"odinger (or Sturm-Liouville) equations, whose analysis
is a complicated task. However several useful informations about
the negative modes of the solutions can be obtained by arguments 
based of catastrophe theory and the inspection of bifurcations
and/or cusps occuring in the pattern of the solutions \cite{Kusmartsev:1989nc, Kusmartsev:cr}.

It is known that the $p$-node BM solution is plagued with $2p$
unstable modes \cite{george}, in other words it is a sphaleron for $p>0$.
As a consequence it is very likely that, at least for small
values of $\phi(0)$, the non-abelian boson star obtained by
deformation of the $p$-th BM solution also posseses $2p$ unstable modes.
(Note also that we could not find
non-trivial  $p=0$ solutions.)
This is at least consistent with the stability of the boson star,
i.e. for $p = 0$.

According to the arguments of the catastrophe theory, we have
to look for bifurcations or the occurence of cusps in the plot
of e.g. the energy (or better the binding
energy given by the difference 
$M-\alpha N$) as a function of the particle number $N$.
Here the main difference between ``pure'' boson stars and the Yang-Mills
boson stars resides in the fact that for the latter the mass of the
solution
does not tend to zero in the limit $N \to 0$ because it converges to
the energy of the corresponding BM solution. As a consequence,
the function $M-\alpha N$ is strictly positive for $N <<1$ suggesting that
the binding energy is positive and conforting the statement that
the $p>1$ Yang-Mills boson stars are indeed unstable.

In the case $p=1$ 
such a diagram indicates the occurence of a
cusp for the maximal value of the particle number
, i.e. at $N \approx 0.35$,
corresponding to
$\phi(0) \approx 0.4$ (the results here correspond to $\alpha=1$).
No solution seems to exist for $N > 0.35$,
however, the solutions existing for higher values of
$\phi(0)$, i.e. $\phi(0) > 0.4$,  form another branch with
a cusp at $N\approx 0.35$.
The  binding energy of the second branch is
higher than the one corresponding to the main branch.
So it is likely that the solution constructed has two unstable
modes for the first branch, i.e. for $\phi(0)<0.4$ and three
unstable modes for $\phi(0)>0.4$.
These results definitely need to
be confirmed by a normal mode analysis of the linearized equations.
\\
\\
{\bf\large Acknowledgements} \\
ER thanks D. H. Tchrakian for useful discussions.
YB is grateful to the
Belgian FNRS for financial support.
The work of ER is carried out
in the framework of Enterprise--Ireland Basic Science Research Project
SC/2003/390 of Enterprise-Ireland.
%
%
%
%
%


\end{document}